\documentclass[twocolumn,secnumarabic,amssymb,nobibnotes,aps,pre]{revtex4}

\newcommand{\del}[1]{\partial_{#1}}
\newcommand{\hf}{\frac{1}{2}}
\newcommand{\intt}{\int_0^\infty}
\newcommand{\grad}{\vec{\nabla}}
\renewcommand{\vec}[1]{\underline{#1}}

\usepackage{graphicx} 
\usepackage{amsmath}

\begin{document}
	
\title{Solvable model of a many-filament Brownian ratchet}

\author{Anthony J. Wood, Richard A. Blythe, Martin R. Evans}

\address{SUPA, School of Physics and Astronomy, University of Edinburgh, Peter Guthrie Tait Road, Edinburgh EH9 3FD}

\begin{abstract}
We construct and exactly solve a model of an extended Brownian ratchet. The model comprises an arbitrary number of heterogeneous, growing and shrinking filaments which together move a rigid membrane by a ratchet mechanism. The model draws parallels with the dynamics of actin filament networks at the leading edge of the cell. In the model, the filaments grow and contract stochastically. The model also includes forces which derive from a potential dependent on the separation between the filaments and the membrane. These forces serve to attract the filaments to the membrane or generate a surface tension that prevents the filaments from dispersing. We derive an $N$-dimensional diffusion equation for the $N$ filament-membrane separations,  which allows the steady-state probability distribution function to be calculated exactly under certain conditions. These conditions are fulfilled by the physically relevant cases of linear and quadratic interaction potentials. The exact solution of the diffusion equation furnishes expressions for the average velocity of the membrane and critical system parameters for which the system stalls and has zero net velocity. In the case of a restoring force, the membrane velocity grows as the square root of the force constant, whereas it decreases once a surface tension is introduced.
\end{abstract}

\maketitle
\section{Introduction}

The Brownian ratchet models a physical system comprising a ratchet-and-pawl device in a surrounding medium \cite{smoluchowski1918ratchet,feynman2011feynman}. Its theoretical interest stems from it providing a mechanism to move a fluctuating object without directly exerting a force on it. Rather it is thermal fluctuations and steric interactions that generate the motion \cite{magnasco1993forced,bang2018experimental} in a manner that is consistent with the second law of thermodynamics. In mathematical terms, the standard Brownian ratchet may be formulated as a drift-diffusive problem for a single spatial co-ordinate \cite{peskin1993cellular}. More recently many-filament systems involving several spatial co-ordinates have been introduced and studied \cite{cole2011brownian, perilli2018force, valiyakath2018polymerisation, whitehouse2018width, mogilner2003force, carlsson2001growth, carlsson2008mathematical, sadhu2018actin, sadhu2019actin, hansda2014branching, wang2014load, das2014collective, tsekouras2011condensation}.

One possible natural manifestation of a many-component ratchet mechanism may be at the boundaries of eukaryotic cells where several actin filaments interact with a restraining cell membrane. Specifically, a network of \emph{actin filaments} grows and contracts in order to move and morph the leading edge of cells \cite{lauffenburger1996cell,svitkina2018actin, andorfer2019isolated}.  The rate of growth of the network is moderated by, among other factors, surrounding monomer concentration \cite{insall2009actin, pujol2012impact, kawska2012actin}. One end of the actin filament (the \emph{barbed} end) elongates at a much higher rate than the other (the \emph{pointed} end), associating a directionality to the growth \cite{pollard1986rate,small1978polarity}. Consequently, the network appears to `treadmill' in one direction with filaments dissociating on the trailing edge \cite{pollard2003cellular}. For the bulk movement of a leading edge (\emph{lamellipodia}), this network tends to be crosslinked, improving the rigidity of the network \cite{lauffenburger1996cell,svitkina1997analysis,matsudaira1994actin}. There are also individual `spikes' out of the cell (\emph{filopodia}), in which the interior actin filaments form a parallel bundle \cite{o1993accumulation,mattila2008filopodia}.

\begin{figure}[tb]
	\centering
	\includegraphics[trim=0mm 0mm 0mm 2mm, clip]{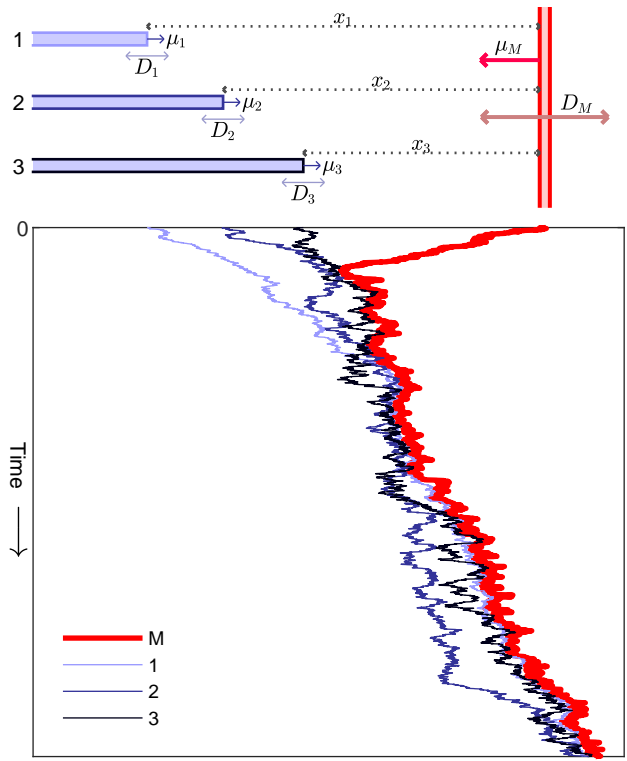}
	\caption{(Colour online) Visualisation and simulation of a continuum ratchet system with $N=3$ filaments. Top: diagram of a three-filament system. Each filament (blue, horizontal) may have a different diffusion constant $D_n$ and drift $\mu_n$. Bottom: realisation of the system over time. The membrane (red, vertical) tends to drift left in isolation, but in the presence of steric interactions with filaments ratchets to the right, settling into a steady state.}
	\label{fig:3FilamentExample}
\end{figure}

In this work, we introduce a general model of an array of $N$ growing and shrinking filaments,  constrained by a rigid drift-diffusing membrane (see Figures \ref{fig:3FilamentExample} and \ref{fig:MainRatchetDiagram}). The model incorporates three major extensions: (i) the filaments are heterogeneous, each characterised by its own polymerisation velocity and variance; (ii) the filaments move under an effective potential with respect to the constraining membrane; (iii) the filaments have long-range, lateral interactions with neighbouring filaments. In this work we consider the case where interactions are attractive, that is, the filaments are attracted to the membrane and/or to each other. The model exhibits the felicitous property of an exactly-solvable steady state, for many parameter choices that correspond to a zero-flux condition that we set out in detail below.

The model that we set out here falls into a class that we refer to as \emph{pure} ratchets \cite{sadhu2019actin}. The defining property of these ratchets is that the membrane moves under thermal fluctuations, and the network grows quickly to occupy any space left vacant (see \cite{peskin1993cellular, cole2011brownian, perilli2018force, valiyakath2018polymerisation, whitehouse2018width} for examples). The key phenomenon that can arise from these pure ratchets, then, is that a membrane that has a natural drift in one direction, may have a net movement in the \emph{opposite} direction, arising exclusively from steric interactions and thermal fluctuations. This is to be distinguished from other systems where filaments directly exert a force on contact and do work to move the membrane \cite{mogilner2003force, carlsson2001growth, carlsson2008mathematical, sadhu2018actin, sadhu2019actin, hansda2014branching, wang2014load, das2014collective, tsekouras2011condensation}. As noted above, the microscopic dynamics of a filament network, involving for example  treadmilling, crosslinking and heterogeneity, is complex \cite{blanchoin2014actin,svitkina2018actin, lauffenburger1996cell,mattila2008filopodia, gardel2004elastic, lieleg2010structure, wear2000actin, insall2009actin}. We do not attempt to model microscopic dynamics in specific detail but instead consider generic  heterogeneous filaments, along with  filament-membrane and interfilament interaction potentials,  which could effectively  encapsulate the dynamical complexity. Specifically, we consider  potentials that serve to attract a filament to the membrane, but does not contribute directly to the membrane motion itself. This  is  a  coarse-grained, effective description of more complicated biological, microscopic effects which may force the filament network to evolve within the locality of the membrane, allowing us to interpret the system as a nonequilibrium steady state. 

In all the  studies discussed so far, a key observable of interest is the steady-state velocity of the membrane. One wishes to understand how the velocity varies with the dynamical properties of the filaments, membrane, and interactions between them. With the model introduced here, we are able to gain exact insight into how the various physical properties of the filaments affect the ability of the overall network to move the fluctuating  membrane. We show how the membrane velocity increases with an increasing harmonic attraction of filaments  to the membrane, but decreases on introducing a surface tension that pulls neighbouring filaments towards one another. The velocity also increases on increasing the diffusion constant of the membrane.

This paper is organised as follows. In Section \ref{sec:ContLim}, we introduce and motivate our system by taking the continuum limit of a \emph{lattice} Brownian ratchet \cite{carlsson2008mathematical,peskin1993cellular}. We then solve for the pdf and membrane velocity, first in Section \ref{sec:ConstDrift} for the case where the filaments have a constant drift, and then in Section \ref{sec:QuadDrift} for where there are effective quadratic interaction potentials. In particular in Section~\ref{sec:restforce} we consider a restoring force towards the membrane and in Section~\ref{sec:surftens} we consider  surface tension across the filament bundle leading edge. We summarise in Section \ref{sec:Summary}.

\section{Model derivation}
\label{sec:ContLim}

\begin{figure}[b]
	\centering
	\includegraphics[trim = 10mm 0mm 0mm 0mm]{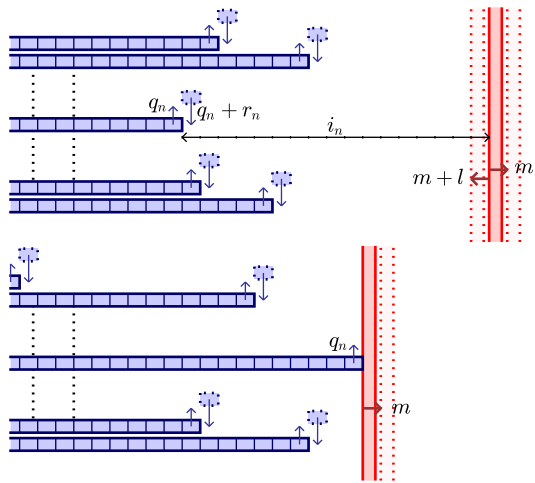}
	\caption{(Colour online) The \emph{lattice} Brownian ratchet model, which is the starting point of the continuum model we solve in this paper. On a lattice, each of the $N$ filaments (blue, horizontal) polymerise and depolymerise, at the rates shown. The membrane (red, vertical) also makes jumps left and right. $i_n \geq 0$ is the integer displacement between filament $n$ and the membrane. In the event of a filament touching the membrane (bottom), the membrane may only move right at the usual rate, and the filament in contact may only contract at its usual rate. The dynamics of the other filaments are unaffected.}
	\label{fig:LatticeRatchet}
\end{figure}
\subsection{Lattice model}
Our starting point is a lattice model of a Brownian ratchet in continuous time, where the discrete lattice represents discretised monomers of the filament.  The reason for starting with a lattice model is that the boundary conditions on the filaments arising from the hard-core exclusion between the filaments and the membrane arise more naturally within the discrete formulation than if one uses a continuum description at the outset.

The dynamics of this lattice model are as follows (see also Figure~\ref{fig:LatticeRatchet} in Appendix~\ref{app:ContLim}). The (rigid) membrane makes unit steps to the left and right at rates defined as $m+l$ and $m$ respectively. Similarly, filament $n$ shrinks (depolymerises) and grows (polymerises) across unit steps at rates $q_n$ and $q_n + r_n$ respectively. Movement is only permitted when a hard-core exclusion interaction is satisfied: the membrane must stay to the right of the right-most filament(s). Thus, the system exhibits ratcheting, where the membrane moves at a velocity different to its inherent drift --- perhaps in the opposite direction entirely --- as a result of thermal fluctuations and steric interactions. The polymer filaments to not exert a force on contact with the membrane, or vice versa. The rate $r_n$ represents the speed of the filament growth and may depend upon the displacement of the filament from the membrane.

\begin{figure*}[t!]
	\centering
	\includegraphics{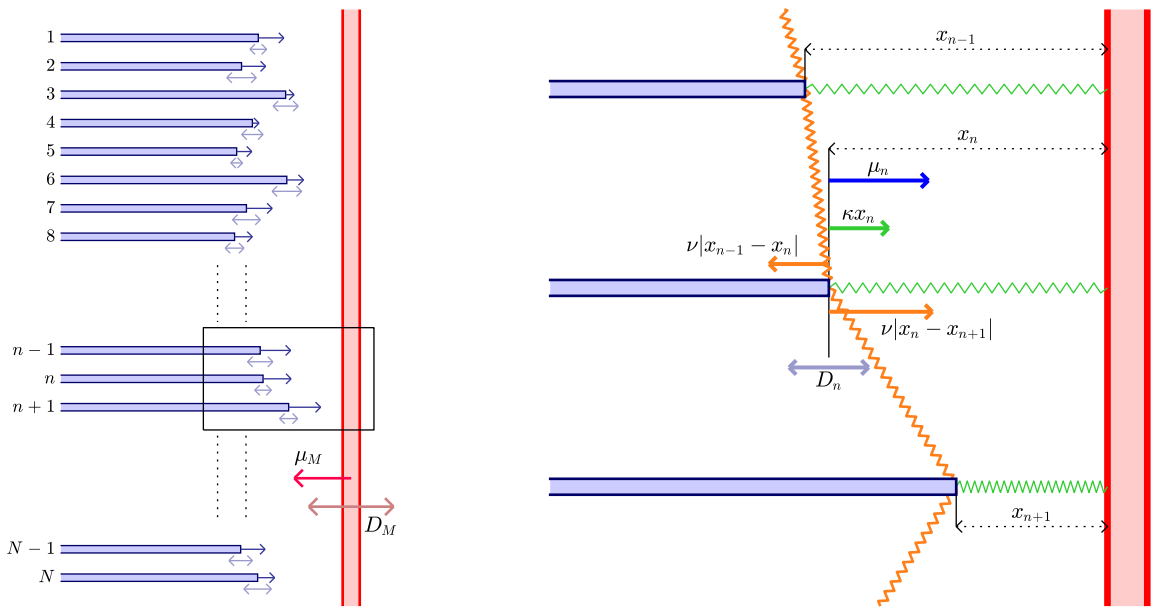}
	\caption{(Colour online)  The complete continuum Brownian ratchet system that we address in this paper. Each of the filaments (blue, horizontal) are growing and shrinking by a diffusion process with coefficient $D_n$, and drift $\mu_n$. The membrane (red, vertical) moves with diffusion coefficient $D_M$ and drift $\mu_M$ towards the filaments. $x_n$ is the displacement between filament $n$ and the membrane. Filament $n$ may then also be attracted to the membrane by a spring force with strength $\kappa$, and also may have a surface tension-like interaction with neighbouring filaments with strength $\nu$.}
	\label{fig:MainRatchetDiagram}
\end{figure*}
Assume now that the system has settled into a steady state, in which the displacements between the filaments and the membrane have stationary distributions. We define $\vec{i} = (i_1,i_2, \dots i_N)$, $i_n \geq 0$ as a vector of displacements between each of the $N$ filaments and the membrane. From here on we treat these displacements $\vec{i}$ as the system \emph{configuration}, although the whole system will in general be drifting at a nonzero velocity (unless it is in a \emph{stalled} state). Define $P_{\vec{i}}$ as the stationary probability of observing the system with displacements $\vec{i}$ under the steady-state condition $\del{t}P_{\vec{i}} = 0$. By considering all possible ways the system can enter and leave configuration $\vec{i}$ (first assuming $i_n > 0$ $\forall n$, so no filaments are in contact with the membrane), the master equation whose solution gives the stationary distribution is
\begin{multline}
\label{eq:LatticeBulkRR}
0 =  -\left[2m+l+\sum_{n=1}^N(2q_n+r_n)\right]P_{\vec{i}} + mP_{\vec{i}-\vec{1}} \\
{} + (m+l)P_{\vec{i}+\vec{1}} + \sum_{n=1}^N \left[q_n P_{\vec{i}-\vec{\hat{n}}}+(q_n+r_n)P_{\vec{i}+\vec{\hat{n}}}\right] \;. 
\end{multline}
Here $\vec{\hat{n}}$ is defined as the unit vector along component $n$, and $\vec{1} \equiv \sum_{n=1}^N \vec{\hat{n}}$.

We now consider the case where filament $k$ makes contact with the membrane and $i_k = 0$, $i_{n\neq k} > 0$. The membrane can now only move to the right, and filament $k$ can only move to the left. In this case, the master equation reads
\begin{multline}
\label{eq:LatticeBoundaryRR}
0 = -\left[m+q_k+\sum_{\substack{n=1 \\ n\neq k}}^N(2q_n+r_n)\right]P_{\vec{i}} + (m+l)P_{\vec{i}+\vec{1}}  \\ 
{} + \sum_{\substack{n=1 \\ n \neq k}}^Nq_nP_{\vec{i}-\vec{\hat{n}}} + \sum_{n=1}^N (q_n+r_n)P_{\vec{i}+\vec{\hat{n}}} 
\end{multline}
for any $k = 1, 2, \dots N$. It is the continuum limit of this equation that furnishes the appropriate boundary condition for the diffusion equation we are about to derive.

\subsection{Continuum limit and diffusion equation}
We now take the limit in which the length of each filament, as well as the position of the membrane, is treated as a continuous random variable. Note that it is in the direction perpendicular to the membrane that the continuum limit is taken; the number of filaments remains discrete (and fixed). We introduce an explicit lattice spacing $a$ such that $\vec{x} = (x_1,x_2,\dots x_N) = a\vec{i}$. The continuum limit then arises by taking $a$ to be small, and then expanding \eqref{eq:LatticeBulkRR} to second order in $a$. In this limit, the probability approaches a pdf that we denote $P(\vec{x})$. From \eqref{eq:LatticeBulkRR}, we then derive a drift-diffusion equation and from \eqref{eq:LatticeBoundaryRR} a set of $N$ boundary conditions. The resulting continuous space system is illustrated in Figure \ref{fig:MainRatchetDiagram}. 

The details of this continuum limit are given in Appendix \ref{app:ContLim}. Here we emphasise the important parameters that emerge. These are the drift and diffusion rates for the membrane (subscript $M$)
\begin{align}
\label{eq:FilamentCLDefinitions}
\mu_M \equiv l \;, \qquad \vec{\mu}_M \equiv \sum_{n=1}^N \mu_M\vec{\hat{n}} \;, \qquad D_M\equiv am \;
\end{align}
and their counterparts for each filament
\begin{align}
\label{eq:MembraneCLDefinitions}
\del{n}V(\vec{x}) \equiv r_n , \qquad D_n\equiv aq_n \;.
\end{align}
In \eqref{eq:MembraneCLDefinitions}  $\del{n} \equiv \del{} / \del{}x_n$
and  the biases (or drifts) $r_n$  derive from a potential $V(\vec{x})$. Note that, as is usual when obtaining a drift-diffusion equation from a lattice-based model,  the diffusion coefficients $D_M$, $D_n$ in \eqref{eq:FilamentCLDefinitions} and \eqref{eq:MembraneCLDefinitions} scale with the lattice spacing. 

We can now express the diffusion equation in terms of the quantities established in (\ref{eq:FilamentCLDefinitions}) and (\ref{eq:MembraneCLDefinitions}) as 
\begin{widetext}
\begin{equation}
\label{eq:DiffusionEquationBulk}
0 = \sum_{n=1}^N\del{n}\left(\del{n}V(\vec{x})+\mu_M + D_M \sum_{k=1}^N \del{k}+ D_n\del{n}\right)P(\vec{x})
\end{equation}
\end{widetext}
and from \eqref{eq:LatticeBoundaryRR} a set of boundary conditions 
\begin{align}
\label{eq:DiffusionEquationBoundary}
0 = \left[\bigg(\del{n}V(\vec{x}) + \mu_M  + D_M\sum_{k=1}^N  \del{k}  +D_n\del{n} \bigg)P(\vec{x})\right]_{x_n=0}.
\end{align}
We refer to (\ref{eq:DiffusionEquationBoundary}) as \emph{zero-current} boundary conditions, because the equation fixes  the probability current at the  boundaries to be zero. To see this, note that the stationary diffusion equation \eqref{eq:DiffusionEquationBulk} can be written as $0 = \grad \cdot \vec{J}$ where $\vec{J}$ is the $N$-component probability current vector and the $n^{\mathrm{th}}$ component of the operator $\grad$ is $\partial_n$. Then \eqref{eq:DiffusionEquationBoundary} is the condition that the $n^{\mathrm{th}}$ component of the current $J_n$ is zero at the boundary $x_n=0$.

The bulk equation \eqref{eq:DiffusionEquationBulk} and boundary conditions \eqref{eq:DiffusionEquationBoundary} fully determine the stationary distribution of filament displacements in our model. We observe that the displacements $\vec{x} = (x_1, x_2, \dots x_N)$ evolve  as a correlated $N$-dimensional diffusion with negative drift. The diffusion of the shared membrane couples the different $x_n$. 

We now highlight the key property of the steady-state equations, \eqref{eq:DiffusionEquationBulk} and \eqref{eq:DiffusionEquationBoundary}, that makes this system exactly solvable under certain conditions. The boundary condition \eqref{eq:DiffusionEquationBoundary} holds at $x_n = 0$. However, if \eqref{eq:DiffusionEquationBoundary} were to hold not just at the boundary but also into the bulk, $x_n \geq 0$, then \eqref{eq:DiffusionEquationBulk} would also be satisfied. In scenarios where this occurs, we can reduce the problem to a set of first order equations that satisfy both equations. We note that for the more general problem of \emph{reflected Brownian motion} with general boundary interactions, closed-form pdfs are not known \cite{franceschi2017explicit, franceschi2017asymptotic}. Therefore the assumption that \eqref{eq:DiffusionEquationBulk} holds in the bulk $x_n \geq 0$, that is that the stationary solution has a zero current everywhere, should be thought of as an ansatz. In a one-filament system this is necessarily the case, however in a higher dimensional system it is possible to have solutions that only have zero current at the boundaries. We will therefore find there are certain restrictions on model parameters that are consistent with the zero-current ansatz. The fact that some particular parameter combinations satisfy this ansatz and some do not is interesting; the systems that do not satisfy this ansatz must contain circulatory currents of probability through the bulk, which one would expect yields a more complex steady state distribution.

For notational convenience, it is helpful to rewrite the zero-current condition \eqref{eq:DiffusionEquationBoundary}, which is now  taken to hold in the bulk, in the vector form
\begin{align}
\left(\grad V(\vec{x})+\vec{\mu}_M + S\grad \right)P(\vec{x})=0
\label{diffansatz}
\end{align} 
where $\vec{\mu}_M$ is specified in Eq.~\eqref{eq:FilamentCLDefinitions} and 
\begin{align}
S = \left(\begin{array}{ccccc} 
D_M+D_1 &D_M & \dots &D_M \\
D_M &D_M+D_2 & \dots &D_M \\
\vdots & \vdots & \ddots & \vdots \\
D_M &D_M & \dots &D_M + D_N 
\end{array}\right) 
\end{align}
is the \emph{diffusion matrix} of the system.  This multi-dimensional  diffusive process then has a drift vector $-\grad V(\vec{x})- \vec{\mu}_M$.

\subsection{Membrane velocity formula}
We now require an expression for the mean membrane velocity, $v_M$, in the steady state. By convention, we take this to be  positive if the membrane is moving to the right. As previously, this is most straightforwardly obtained within the lattice model, so we write down the lattice version first and then take the continuum limit. This is detailed in Appendix~\ref{app:ContLim}, and we obtain
\begin{equation}
v_M =  
-\mu_M + D_M\sum_{n=1}^N\left[\prod_{\substack{m=1 \\ m\neq n}}^N\left(\intt dx_m\right)P(\vec{x}|_{x_n = 0})\right] \label{eq:MembraneVelocityExpression}
\end{equation}
where $P(\vec{x}|_{x_n = 0})$ is the pdf evaluated at $x_n=0$. This equation has an intuitive form: the membrane tends to move left at speed $\mu_M$, but is then biased right by an amount that increases with increasing contact of the membrane with filaments. We note that $v_M$ can take either sign: the membrane can move in either direction. If $v_M=0$ the system has \emph{stalled}.

\subsection{Introductory example: single filament} \label{sec:SingleFilamentExample}

As a familiarisation exercise, we first solve the model in the case of a single filament. The filament grows and contracts stochastically, with a constant drift $\mu_1$ towards the membrane along with a restoring force $\kappa x_1$ and diffusion constant $D_1$. The membrane has a diffusion constant $D_M$, and a drift $\mu_M$ towards the filament. We stress that there is an asymmetry in this interaction: the restoring force $\kappa x_1$ attracts the filament to the membrane, but not vice versa. This is equivalent to a one-dimensional drift-diffusion, in a harmonic potential and a reflecting boundary at zero \cite{schultenlectures}.

For a single filament the zero current boundary condition implies that (\ref{diffansatz}) must hold for all $x_1$ and the condition reads
\begin{align}
0 = \left[\kappa x_1 + \mu_1 + \mu_M + (D_1+D_M)\del{1}\right]P(x_1)\;.
\end{align}
 This is straightforwardly integrated to give
\begin{align}
P(x_1) = \mathcal{A}^{-1} \exp\left(-\frac{\hf \kappa x_1^2 + (\mu_1+\mu_M)x_1}{D_1+D_M}\right) \;.
\end{align}
The normalisation $\mathcal{A}$ is fixed by the condition $\intt dx_1 P(x_1)=1$ which yields
\begin{align}
\mathcal{A} & = \sqrt{\pi} \sqrt{\frac{D_1+D_M}{2\kappa }} {\rm e}^{c^2} \text{erfc}\left(c\right)
\end{align}
where $\mbox{erfc}(\alpha) = 2/\sqrt{\pi}\int_\alpha^\infty e^{-t^2}dt$ is the complimentary error function, and $c = (\mu_1+\mu_M)/\sqrt{2\kappa(D_1+D_M)}$. With this, we find using \eqref{eq:MembraneVelocityExpression} the membrane velocity
\begin{align}
v_M & = -\mu_M + D_M P(0) \\
& = -\mu_M + \frac{D_M\sqrt{\frac{2\kappa}{D_1+D_M}} \exp\left({-\frac{\left(\mu_1+\mu_M\right){}^2}{2 \kappa  (D_1+D_M)}}\right)}{\sqrt{\pi}\;\text{erfc}\left(\frac{\mu _1+\mu_M}{\sqrt{2\kappa(D_1+D_M)}}\right)} \;. \label{eq:1DSpringVelocity}
\end{align}

\begin{figure}[tb]
	\centering
	\includegraphics[trim=0mm 0mm 0mm 0mm]{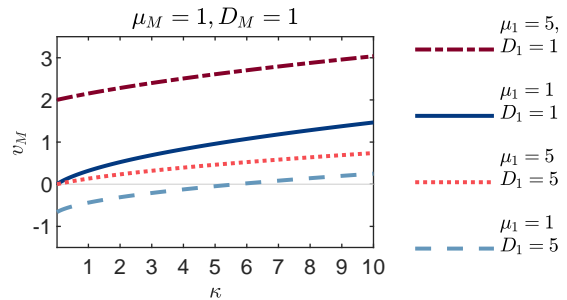}
	\caption{(Colour online) Analytic membrane velocity $v_M$ \eqref{eq:1DSpringVelocity} for a single filament system, for four different filaments. $v_M$ is a monotonically increasing function of the restoring force constant $\kappa$. Against the same membrane ($\mu_M = 1$, $D_M = 1$), we see that less diffusive, higher drift filaments are more effective at moving the membrane.}
	\label{fig:SingleFilamentVelocity}
\end{figure}

We plot $v_M$ for various filaments in Figure \ref{fig:SingleFilamentVelocity}. $v_M$ is a monotonically increasing function of $\kappa$. For the example $\mu_1 = 1$, $D_1 = 5$ (red, dashed), we see that the membrane can have a positive, negative or zero velocity depending on the value of $\kappa$. Thus a large enough restoring force will always lead to a positive velocity. In the case $\mu_M + \mu_1 = 0$, for which the filament and membrane a relative drift  towards each other only due to the linear restoring $\kappa x_1$,  Equation~\eqref{eq:1DSpringVelocity} reduces to
\begin{align}
\label{eq:SingleFilamentVelocityNoDrift}
v_M = -\mu_M + \frac{D_M}{\sqrt{\pi}}\sqrt{\frac{2\kappa}{D_1+D_M}}
\end{align}
and the velocity deviates from the free velocity $-\mu_M$ as the square root of the force constant $\kappa$. We show in Section \ref{SSec:AnalyticKappaScaling} that this scaling holds for $N$ filaments.

\section{Constant drift solution for many filaments} \label{sec:ConstDrift}

We now solve the system for $N$ filaments. First, we consider the case of a linear potential $V(\vec{x})$,  implying constant drifts for each filament. That is,
\begin{align}
\qquad V(\vec{x}) = \vec{\mu}_F\cdot\vec{x}\;, \qquad \vec{\mu}_F \equiv \sum_{n=1}^N \mu_n\vec{\hat{n}}
\end{align}
with the subscript $F$ denoting the filaments.  The zero-current condition \eqref{diffansatz} now reads
\begin{align}
\label{eq:DiffusionEquationBoundary2}
\left(\vec{\mu}_M+\vec{\mu}_F + S\grad \right)P(\vec{x}) = 0 \;.
\end{align}
To satisfy this condition let us assume a normalised, trial solution
\begin{align}
\label{eq:LinearTrialSol} 
P(\vec{x}) = \left(\textstyle \prod_{n=1}^N \lambda_n\right)e^{-\vec{\lambda}\cdot\vec{x}}
\end{align}
with $\vec{\lambda} = (\lambda_1, \lambda_2, \dots \lambda_N)$. This solution has exponential decay of the filament-membrane separations with decay constants $\lambda_n$ and the distributions for individual filaments are decoupled, despite the fluctuating  membrane coupling the $x_n$ to one another. Substituting this trial solution into \eqref{eq:DiffusionEquationBoundary2} leads to the constraint
\begin{align}
\vec{\mu}_M + \vec{\mu}_F - S \vec{\lambda} = 0
\end{align}
which in turn implies
\begin{align}
\vec{\lambda} = S^{-1}\left(\vec{\mu}_M+\vec{\mu}_F\right) \;.
\end{align}
Furthermore, the entries of $S^{-1}$ are explicitly calculable for any $N$ via the Sherman-Morrison formula \cite{bartlett1951inverse}:
\begin{align}
\label{eq:InverseS}
(S^{-1})_{nk} = D_n^{-1}\left(\delta_{nk} - \frac{D_k^{-1}}{D_M^{-1}+\sum_{n'=1}^ND_{n'}^{-1}}\right) \;. 
\end{align}
With further algebra, the components of $\vec{\lambda}$ reduce to
\begin{align}
\label{eq:LambdaExpression}
\lambda_n = D_n^{-1}\left(\mu_n + \frac{\mu_MD_M^{-1}-\sum_{n'=1}^N\mu_{n'}D_{n'}^{-1}}{D_M^{-1}+\sum_{n''=1}^ND_{n''}^{-1}}\right)
\end{align}
giving an explicit solution for $P(\vec{x})$ as a function of the diffusion and drift parameters of the system. We see that $\lambda_n$, the exponential decay constant  for the separation, increases with drift $\mu_n$ but decreases with diffusion constant $D_n$. However the dependence on the drift and diffusion constants of the other filaments appears rather complicated. We shall  see that the interdependencies are best  understood  when we consider the membrane velocity.

\subsection{Mean membrane velocity}

We initially consider the case where all $\lambda_n > 0$ (see Section \ref{SSec:SteadyStateCondition} for discussion of when this does not hold). With the decoupled exponential form \eqref{eq:LinearTrialSol} of $P(\vec{x})$, the membrane velocity \eqref{eq:MembraneVelocityExpression} is straightforward to calculate as
\begin{align}
v_M &= -\mu_M +D_M\sum_{n=1}^N \lambda_n \\
& = \frac{-\mu_MD_M^{-1}+\sum_{n=1}^N\mu_nD_n^{-1}}{D_M^{-1}+\sum_{n'=1}^ND_{n'}^{-1}} \;. \label{eq:MembraneVelocity} 
\end{align}
Equation  \eqref{eq:MembraneVelocity} is the central result of this section and gives the membrane velocity in terms of all the  constituent filament drift and diffusion constants  $\{ \mu_n,D_n \}$.

The exponential decay constants $\lambda_n$ can then be written
\begin{equation}
\lambda_n = \frac{\mu_n-v_M}{D_n}
\end{equation}
with the numerator of $\lambda_n$ being the difference between the drift of filament $n$ and the net velocity of the membrane determined by the whole system. As this difference decreases, the average separation $\langle x_n \rangle = \lambda_n^{-1}$ increases.

The membrane stalling  drift $\mu_M^*$ is defined as the drift for which $v_M = 0$:
\begin{align}
\mu_M^* = D_M\sum_{n=1}^N \frac{\mu_n}{D_n}\;.
\label{stallforce}
\end{align}
This result can be interpreted in terms of the ratcheting mechanism. $\mu_M^*$ increases  as the drift of each filament $\mu_n$ increases. Thus the membrane must have large drift to the left to stall the ratchet mechanism arising from more strongly polymerising filaments.
However $\mu_M^*$ decreases as each $D_n$ increases. Thus greater variability of the polymerisation process reduces  any ratcheting effect. On the other hand, increasing the membrane diffusion constant $D_M$ increases $v_M$ and thus  requires an increase in membrane drift to stall the ratchet mechanism. This is because the fluctuations in membrane position due to a large  $D_M$  afford more opportunity for polymerisation near the  membrane.

\subsection{Steady-state condition} \label{SSec:SteadyStateCondition}

A property of the membrane-filament system is that it may not reach a steady state. If at least one of the $\lambda_n$ is negative, then \eqref{eq:LinearTrialSol} is not normalisable, indicating the absence of a steady state. Physically, this arises from one or more of the filaments drifting away from the membrane in perpetuity. Thus the requirement for a steady state in which the filaments travel with the membrane is that $\lambda_n > 0$ for all $n = 1, 2, \dots N$.

To determine when this requirement holds, we first note from \eqref{eq:LambdaExpression} that the sign of each $\lambda_n$ is dependent on each and every other filament. Given $N$ filaments with a set of parameters $\{D_n,\mu_n\}$ and a membrane with a given $\mu_M$, $D_M$, we then need to determine whether the full system forms a steady state.

Label the filaments $1, 2 \dots N$ in order of decreasing drift, such that $\mu_1 \geq \mu_2 \geq \dots \geq \mu_N$. We first check if the   filament with the highest drift ($\mu_1$) would form a steady state with the membrane, if it were the only filament in the system. From the form of $\lambda_n$ for $N=1$, this gives the trivial condition $\mu_1 + \mu_M > 0$. If this is satisfied, filament $1$ participates in the steady state because it moves towards the membrane. If it does not, the membrane and the filament drift apart, and no steady state is formed. Furthermore, as $\mu_1 \geq \mu_2 \dots \geq \mu_N$, \emph{none} of the filaments settle into a steady state. 

We now add filament $2$. We check if $\lambda_2 > 0$. From the form of $\lambda_n$ for $N=2$, this gives the condition $\mu_2 > \left(-\mu_MD_M^{-1}+\mu_1D_1^{-1}\right)/\left(D_1^{-1}+D_M^{-1}\right)$. If this is satisfied, filament $2$ participates in the steady state. If it is not, the one filament-membrane system runs away from filament $2$, and also the remaining filaments.

We repeat this process sequentially, and assuming that the condition has been satisfied by all filaments up to $n-1$, we add filament $n$. The requirement for $\lambda_n > 0$ is 
\begin{equation}
\mu_n > \frac{-\mu_MD_M^{-1}+\sum_{n'=1}^{n-1}\mu_{n'}D_{n'}^{-1}}{D_M^{-1}+\sum_{n''=1}^{n-1}D_{n''}^{-1}} \;.
\end{equation}
We find a result that, in retrospect, is self-consistent and physically intuitive: filament $n$ will participate in the steady state if $\mu_n$ is greater than the steady state membrane velocity \eqref{eq:MembraneVelocity} from the system of the $n-1$ faster filaments. This is independent of $D_n$--- the diffusivity of a filament does not affect whether it can `catch up' with a system in the long term.

\begin{figure}[tb]
	\centering
	\includegraphics[trim = 0mm 0mm 0mm 0mm, clip]{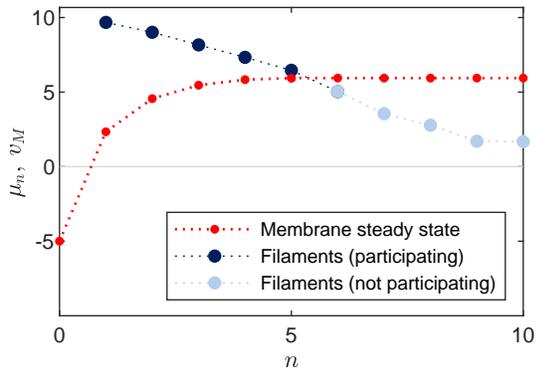}
	\caption{(Colour online) Sequentially adding filaments to a system with a membrane with $\mu_M = 5$, by decreasing velocity. All diffusion parameters are set to $1$ for simplicity. In this example, filament $6$ is slower than the membrane when it is added, so filaments $1$--$4$ form a steady state and other filaments fall away.}
	\label{fig:ContLimspeedplot}
\end{figure}

Each additional participating filament contributes to increasing $v_M$. We must then sequentially add filaments by decreasing velocity, until a filament is found that is slower than $v_M$ up to that point. Then, that filament and all lower velocity filaments do not participate in the steady state, and the pdf $P(\vec{x})$ is constructed from the participating filaments only. This procedure is illustrated in Figure \ref{fig:ContLimspeedplot}, where filaments are sequentially added, and a new $v_M$ is calculated on the addition of each filament.

In the case of a large number of identical filaments $D_1, \dots D_N = D_F$, $\mu_1, \dots \mu_N = \mu_F$, we find
\begin{align}
v_M &= \frac{-\mu_MD_M^{-1}+N\mu_FD_F^{-1}}{D_M^{-1}+ND_F^{-1}} \\
&\approx \mu_F - \frac{1}{N}\frac{D_F}{D_M}\left(\mu_M+\mu_F\right) \;.
\end{align}
We see that the membrane velocity converges to the filament drift $\mu_F$ as the number of filaments $N\to \infty$. This specific case has been previously derived in \cite{cole2011brownian,valiyakath2018polymerisation}. 

\section{Quadratic potential solution} \label{sec:QuadDrift}

Until now, we have considered the case where there are no explicit forces between the filaments or between the filaments and the membrane. We now introduce interactions between components of the system that take the form of linear restoring forces that derive from quadratic interaction potentials. As we now show, this system is also exactly solvable within the zero-current ansatz \eqref{diffansatz} for a subset of all possible interactions of this type.

To this end, we specify a potential consisting of general linear and quadratic terms
\begin{align}
V(\vec{x}) = \vec{\mu}_F\cdot\vec{x} + \hf \vec{x}^T\Gamma\vec{x}\;,
\end{align}
where $\Gamma$ is a symmetric matrix that describes the interaction at quadratic order. 
Each diagonal element of the quadratic term represents a harmonic potential for the separation between a filament and the membrane. The off-diagonal terms represent couplings between the different filaments.

Under this potential, the ansatz \eqref{diffansatz} reads
\begin{align}
\label{eq:RatchetBCVarExp}
\left(\vec{\mu}_M + \vec{\mu}_F + \Gamma\vec{x} + S\grad \right)P(\vec{x})  = 0 \;.
\end{align}
Given this quadratic form of the potential, we choose as a trial solution for \eqref{eq:RatchetBCVarExp} the pdf
\begin{align}
\label{eq:TrialSolutionKappa}
P(\vec{x}) = \mathcal{A}^{-1}e^{-\vec{\lambda}\cdot\vec{x}-\hf\vec{x}^TG\vec{x}} \;.
\end{align}
The exponent in \eqref{eq:TrialSolutionKappa} contains all possible linear and quadratic combinations of the $x_n$. $\mathcal{A}$ is a normalising constant and $G$ is a symmetric matrix.

Inserting this trial solution in \eqref{eq:RatchetBCVarExp} yields
\begin{equation}
\vec{\mu}_M + \vec{\mu}_F+\Gamma\vec{x}-S\left(\vec{\lambda}+G\vec{x}\right) = 0 \;.
\end{equation}
This condition implies a solution for $\vec{\lambda}$
\begin{align}
\label{eq:G}
\vec{\lambda} = S^{-1}\left(\vec{\mu}_M+\vec{\mu}_F\right)\;,\quad\mbox{where} \quad S^{-1}\Gamma =G \;.
\end{align}
As $G$ is symmetric, for \eqref{eq:TrialSolutionKappa} to be a valid solution, we  must have $S^{-1}\Gamma$ symmetric, which is not generally the case. Thus the trial solution \eqref{eq:TrialSolutionKappa}  does not satisfy the ansatz \eqref{diffansatz} in the general case of several filaments. A possible reason for this is that the $x_n = 0$ zero-current conditions \eqref{diffansatz} may not always extend into the bulk. Then, there would be additional probability currents in the bulk and the filament-membrane displacements would form a more complex nonequilibrium steady state.

In light of this, we seek particular systems for which $G = S^{-1}\Gamma$ \emph{is} symmetric. With reference to Figure \ref{fig:MainRatchetDiagram}, we address two cases. First, a system where the filaments are attracted to the membrane by a restoring spring-like force with strength $\kappa$. Then, we introduce an additional surface tension with strength $\nu$.

We note that the pdf \eqref{eq:TrialSolutionKappa} is a multivariate normal distribution \cite{genz2009computation}. As the domain of $P(\vec{x})$ is restricted to the upper \emph{orthant} $x_n \geq 0$, the normalisation factors $\intt dx_1 \dots \intt dx_N P(\vec{x})$ are challenging to evaluate exactly for large $N$ \cite{genz2009computation}. Regardless of this we can still analyse $P(\vec{x})$ and in particular find scaling laws for $v_M$.

\subsection{Restoring force between filaments and the membrane}
\label{sec:restforce}
We can incorporate a harmonic potential with strength $\kappa > 0$. This is by design an asymmetric interaction  with attracts each filament to the membrane, but not vice versa. We hope to encapsulate the features of a larger membrane moving in a viscous medium, and a rapidly evolving network of actins with a variable rate of association and dissociation \cite{svitkina2018actin}.

This interaction is incorporated with the diagonal matrix $\Gamma_{nm} = \kappa \delta_{nm}$. This linear restoring force is intended to model effective interactions between the filaments and membrane. We then find from \eqref{eq:G} that the matrix $G = \kappa S^{-1}$ is symmetric (as required) because $S$ is symmetric --- see \eqref{eq:InverseS}. Then the stationary solution $P(\vec{x})$ is obtained from \eqref{eq:TrialSolutionKappa} as
\begin{align}
P(\vec{x}) =
\mathcal{A}^{-1}\exp\left(-\hf\underline b^T S^{-1}\underline b\right)
\label{eq:TrialSolutionKappa2}
\end{align}
where
\begin{equation}
\underline b = \kappa^\hf\vec{x}+\kappa^{-\hf}\left(\vec{\mu}_M + \vec{\mu}_F\right)\;.
\end{equation}

As each of the filaments is now in a harmonic trap with respect to the membrane, one expects all filaments to participate in the steady state i.e. none lag behind. In other words, \eqref{eq:TrialSolutionKappa2} approaches zero as any of the $x_n \to \infty$. Finally, note that unlike the linear drift case \eqref{eq:LinearTrialSol}, these quadratic potential systems contain combinations of the form $x_nx_m$ in the pdf, implying that the distribution does not decouple over filaments.

\subsubsection{Velocity scaling law} \label{SSec:AnalyticKappaScaling}
We now argue that the introduction of a harmonic interaction introduces a $\kappa^\hf$ enhancement to the membrane velocity. The normalisation constant $\mathcal{A}$ is found by requiring
\begin{align}
\prod_{n=1}^N\left(\intt dx_n\right)  P(\vec{x}) = 1 \;.
\end{align}
After a variable change, this is written 
\begin{align}
\mathcal{A} = \kappa^{-\frac{N}{2}}\prod_{n=1}^N\left(\int^\infty_{\kappa^{-\hf}(\mu_M+\mu_n)} dx'_n\right)  e^{-\hf\vec{x}'^T S^{-1}\vec{x}'} \;.
\end{align}
When $\kappa$ is large, we can approximate the lower bound of each of the $N$ integrals to extract the dominant $\kappa$-dependence
\begin{align}
\mathcal{A} & \approx \kappa^{-\frac{N}{2}}\prod_{n=1}^N\left(\intt dx'_n\right)e^{-\hf\vec{x}'^T S^{-1}\vec{x}'}   \\ 
& \equiv \mathcal{B} \kappa^{-\frac{N}{2}} \label{eq:vmintapprox} \;. 
\end{align}
We define $\mathcal{B}$ as a $\kappa$-independent constant. We repeat this method to extract the $\kappa$-dependence from the $(N-1)$-dimensional integrals in \eqref{eq:MembraneVelocityExpression} to give an overall scaling for the membrane velocity 
\begin{align}
v_M & = -\mu_M + D_M\sum_{n=1}^N\left[\prod_{\substack{m=1\\ m \neq n}}^N\left(\intt dx_m\right)P(\vec{x}|_{x_n = 0})\right]  \\
& \approx -\mu_M + \mathcal{C} \kappa^\hf \;. \label{eq:MembraneVelocityKappa}
\end{align}
\begin{figure}[t]
	\centering
	\includegraphics[trim = 0mm 0mm 0mm 0mm]{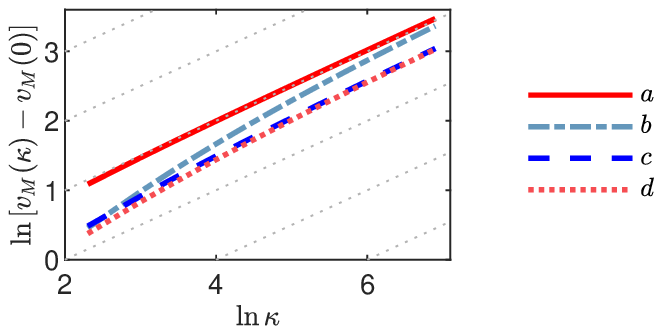}
	\caption{(Colour online)  Numerically integrated membrane velocity as a function of $\kappa$ for four different $N=3$ filament systems $(a)$---$(d)$. For each of these parameter sets, the gradients indicate a $v_M \propto \kappa^\hf $ scaling relationship for large $\kappa$. \\ 
		$(a)$ $D_M = 1$, $\mu_M=-1$, $D_F = 1$, $\vec{\mu}_F = (1,3,1)$, $\nu = 2$, \\
		$(b)$ $D_M = 1$, $\mu_M = 3$, $D_1 = 2$, $D_2 = 1/2$, $D_3 = 3$, $\vec{\mu}_F = (-2,5,2)$, $\nu = 0$, \\
		$(c)$ $D_M = 1/2$, $\mu_M = 2$, $D_1 = 1/2$, $D_2 = 2$, $D_3 = 1$, $\vec{\mu}_F = (1,2,1)$, $\nu = 0$, \\
		$(d)$ $D_M = 1/2$, $\mu_M=5$, $D_F = 1$, $\vec{\mu}_F = (2,-1,-1)$, $\nu = 1$. }
	\label{fig:MembraneVelocityPlotKappa2}
\end{figure}
$\mathcal{C}$ is another $\kappa$-independent constant. We expect the correction to approximation \eqref{eq:vmintapprox} to be of order $\kappa^{-(N+1)/2}$, corresponding to an $\mathcal{O}(\kappa^0)$ correction to \eqref{eq:MembraneVelocityKappa}. To support this, we present in Figure \ref{fig:MembraneVelocityPlotKappa2} the numerically integrated membrane velocities against $\kappa^\hf $ for four $N=3$ filament systems, each with different sets of diffusion and drift parameters. In all four cases we observe a $\kappa^\hf $ scaling for large $\kappa$. In the case $-\mu_M = \mu_1 = \mu_2 = \dots = \mu_N$, the approximations in \eqref{eq:vmintapprox}, \eqref{eq:MembraneVelocityKappa} become exact, as we saw in \eqref{eq:SingleFilamentVelocityNoDrift}.

\subsection{Surface tension}
\label{sec:surftens}

We now add an attractive interaction between neighbouring filaments. Again, we choose the simplest interaction, which is one that derives from a harmonic potential. This serves to equalise the length of neighbouring filaments, and thus models a surface tension in the filament bundle.

This additional interaction leads to a second term appearing in the potential $V(\vec{x})$, 
\begin{align}
V(\vec{x}) = \hf \kappa \sum_{n=1}^N x_n^2 + \hf \nu \sum_{n=1}^{N-1} (x_{n+1}-x_n)^2 \;,
\end{align}
where the parameter $\nu$ specifies the strength of the surface tension. The interaction matrix is then
\begin{align}
\Gamma = \left(\begin{array}{cccccc} 
\kappa+\nu &-\nu & \cdot & &\cdot  &\cdot \\
-\nu &\kappa+2\nu & -\nu & &\cdot  &\cdot \\
\cdot & -\nu & \kappa+2\nu & &\cdot  &\cdot \\
&  &  &\ddots  & &  \\
\cdot & \cdot & \cdot  & & \kappa+2\nu & -\nu \\
\cdot & \cdot & \cdot  & &-\nu & \kappa+\nu 
\end{array}\right) \;.
\end{align}
Note that we have assumed free boundary conditions: that is, filaments $1$ and $N$ each have only a single neighbour.

With this interaction matrix, the matrix $G = S^{-1}\Gamma$ that appears in the stationary solution (\ref{eq:TrialSolutionKappa}) is symmetric only if the $N$ filament diffusivities each take the same value, which we denote $D_F$. Then,
\begin{widetext}
\begin{equation}
G_{nm} = \kappa D_F^{-1}\bigg(\delta_{nm}-\frac{D_F^{-1}}{D_M^{-1}+ND_F^{-1}}\bigg) \\
{}+\nu D_F^{-1}\left(2\delta_{nm}- \delta_{n,m-1}-\delta_{n,m+1}-\delta_{n1}\delta_{m1}-\delta_{nN}\delta_{mN}\right) \;. 
\end{equation}
\end{widetext}
With this form of $G$, \eqref{eq:TrialSolutionKappa} is the pdf for a system with inhomogeneous drift terms, a restoring force to the membrane, and a surface tension. 

\subsubsection{Example: two filaments with quadratic interactions}
To illustrate the previous result, we explicitly calculate the membrane velocity for the $N=2$ filament case, with both quadratic interactions included. For two filaments with $\mu_M = \mu_1 = \mu_2 = 0$, the pdf \eqref{eq:TrialSolutionKappa} becomes explicitly
\begin{multline}
\label{eq:TrialSolutionKappa3}
P(\vec{x}) = \mathcal{A}^{-1}\exp\left(-\frac{\nu  \left(x_1-x_2\right)^2}{2 D_F}\right)\times \\
\exp \left(-\kappa \frac{\left(x_1^2+x_2^2\right) (D_F+D_M)-2 D_M x_1 x_2}{2 D_F (D_F+2 D_M)}\right) \;.
\end{multline}
Here, the filaments move towards the membrane by the restoring force only. In this case, the normalisation constant $\mathcal{A}$, obtained by integrating over all $x_1>0$ and $x_2>0$, has the exact form 
\begin{multline}
\mathcal{A} = \sqrt{\frac{D_F (D_F+2 D_M)}{\kappa  (\kappa +2 \nu )}}\times \\
\left[\tan ^{-1}\left(\frac{D_F \nu +D_M (\kappa +2 \nu )}{\sqrt{\kappa (\kappa +2 \nu )
		D_F(D_F+2 D_M)}}\right)+\frac{\pi}{2} \right] 
\end{multline}
where we have used Eq.~4.3.2 in \cite{ng1969table} to evaluate the integral.
Then, the membrane velocity follows from \eqref{eq:MembraneVelocityExpression}:
\begin{equation}
v_M  = \frac{\sqrt{2 \pi } D_M \sqrt{\frac{\kappa }{\frac{D_F (\kappa +\nu )}{\kappa +2 \nu }+D_M}}}{\tan ^{-1}\left(\frac{D_F \nu +D_M (\kappa +2 \nu )}{\sqrt{\kappa (\kappa +2 \nu )
			D_F(D_F+2 D_M)}}\right)+\frac{\pi}{2} } \;.
\end{equation}

\begin{figure}[tb]
	\centering
	\includegraphics[trim = 0mm 0mm 0mm 0mm]{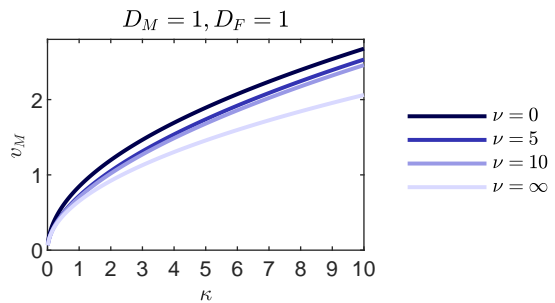}
	\caption{(Colour online) Analytic membrane velocity as a function of $\kappa$, for a two-filament system at four different surface tension strengths $\nu$. On increasing $\nu$ the filaments become less effective at moving the membrane, with the limiting case $\nu \to \infty$ effectively a one-filament system \eqref{eq:MembraneVelocityInfiniteNu}.}
	\label{fig:MembraneVelocityPlotKappa}
\end{figure}
For the case $\nu=0$ (i.e. where there is no surface tension), we find that the velocity is proportional to $\kappa^\hf $, as claimed in the previous subsection. This function is plotted in Figure \ref{fig:MembraneVelocityPlotKappa}. For a fixed $\kappa$, the membrane velocity \emph{decreases} as the surface tension strength increases. The limit of $v_M$ as $\nu \to \infty$ is
\begin{align}
\label{eq:MembraneVelocityInfiniteNu}
\lim_{\nu\to\infty} v_M = \frac{D_M}{\sqrt{\pi }} \sqrt{\frac{2\kappa }{D_F/2+D_M}} \;.
\end{align}
In this limit the two filaments are tightly bound and resemble a single filament \eqref{eq:SingleFilamentVelocityNoDrift}, with diffusion constant $D_F/2$. 

\subsubsection{More than two filaments}

\begin{figure}[tb]
	\centering
	\includegraphics[trim=0mm 3mm 0mm 3mm]{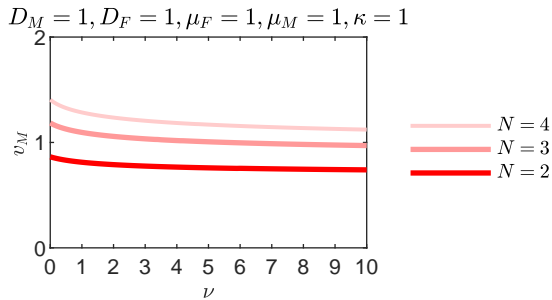}
	\caption{(Colour online) Membrane velocity as a function of surface tension strength $\nu$ for up to $N=4$ filaments, calculated by numerical integration of the pdf \eqref{eq:TrialSolutionKappa}. While increasing $N$ increases $v_M$, the velocity decreases with $\nu$ for all three systems as they become less effective at moving the membrane.}
	\label{fig:SurfaceTensionPlot}
\end{figure}

In the case of more than two filaments, it is difficult to calculate the normalisation constant $\mathcal{A}$ in \eqref{eq:TrialSolutionKappa} in a convenient form. Therefore, to investigate this case, we turn to numerical evaluation of both the normalising integral and the integrals that appear in the expression for the membrane velocity \eqref{eq:MembraneVelocityExpression}. We plot the membrane velocity as a function of surface tension strength for fixed drift and diffusion rates in Figure~\ref{fig:SurfaceTensionPlot}. For all $N = 2,3,4$, we find that the membrane velocity decreases with surface tension, asymptotically approaching a constant. 

There is a straightforward physical interpretation of this result. The ratcheting mechanism means that only  a single filament need be in contact with the membrane in order to force it to move right. By introducing a surface tension, there will always be a force on the closest filament from its neighbours that pulls it \emph{away} from the membrane, making the filament network as a whole less efficient at ratcheting the membrane. 
\section{Summary and outlook}
\label{sec:Summary}

In this work we have derived the steady-state distribution of a pure ratcheting system of $N$ heterogeneous filaments, constricted by a membrane. This model exhibits ratcheting, whereby a membrane moves at a velocity different to its inherent drift, solely due to thermal fluctuations and steric interactions between it and the filaments. This provides a more comprehensive, general formalism than earlier continuum models \cite{cole2011brownian, valiyakath2018polymerisation}. Our solution relies on the zero-current condition  which reduces the drift-diffusion problem  to first order equations. We have found that the zero-current condition  holds for  a variety of  systems including physically relevant cases of fixed filament drift (linear filament-membrane interaction potentials) and quadratic filament-membrane  and quadratic filament-filament interaction potentials.

For these cases, one can find explicit expressions for the distribution of filament displacements (e.g. \eqref{eq:LinearTrialSol} and \eqref{eq:LambdaExpression} for the constant drift case) and from these one can derive expressions for the membrane velocity. In the case of an arbitrary number $N$ of heterogeneous filaments, each with its own fixed drift and diffusion constant, we have obtained an explicit and transparent expression \eqref{eq:MembraneVelocity} for the membrane velocity $v_M$, and in \eqref{eq:MembraneVelocityKappa} a scaling law for when the filaments are also attracted to the membrane by a restoring force. Equation \eqref{eq:MembraneVelocity} reveals {\em inter alia} how the ratcheting mechanism is enhanced by greater membrane diffusion.

For the case of constant-drift filaments, the pdf \eqref{eq:LinearTrialSol} decouples among each of the $N$ filaments. However, a subtlety arises in that it is not obvious as to whether a collection of filaments will actually form a steady state. A new filament will only participate if its velocity is greater than the prior membrane velocity. Conversely, one new high-velocity filament can disrupt a pre-existing steady state, by pulling the system away from other lower velocity filaments. Which filaments participate is a collective outcome of the set of filaments, and may be determined by considering the filaments in decreasing order of drift velocity (Figure \ref{fig:ContLimspeedplot}).

For the case of a linear restoring force, all filaments will participate in the steady state. While it is a challenge to normalise the pdf \eqref{eq:TrialSolutionKappa} for large $N$ we find in \eqref{eq:MembraneVelocityKappa} that a harmonic attraction to the membrane increases the velocity by an amount proportional to the square root of the force constant $\kappa$, to leading order. It is physically intuitive that the velocity would increase as the attractive force increases, however the exponent of $1/2$ in \eqref{eq:MembraneVelocityKappa} is less obvious. 

Finally, we have introduced a surface tension element between neighbouring filaments, and shown that $v_M$ decreases as a result. Intuitively, a surface tension will always pull the right-most filament away from the membrane, giving the membrane more space to freely move left. This suggests that the filament network most efficiently moves the membrane when each filament moves independently of one another. 

An interesting problem that arises from this work is that some particular parameter combinations have zero probability current in the bulk, and some do not. In these non ansatz-satisfying systems, one should expect circulatory --- perhaps oscillatory --- flows of probability current in the bulk. A natural progression from the work presented here would be to further probe these more complex systems, and how the tuning of these parameters gives rise to additional bulk currents.

This system is exactly solvable and the expressions for the membrane velocity $v_M$ are analytic, for an arbitrary number of filaments. In contrast, the discrete case of Figure \ref{fig:LatticeRatchet} does not permit a separable solution. To more closely resemble the dynamics of real actin networks, and to extend beyond the pure ratchet model considered here,  it would be desirable to encode some type of  direct \emph{contact force} between the filaments and membrane beyond hard-core exclusion \cite{ananthakrishnan2007forces}. The challenge is  that for any non-instantaneous contact (such as \emph{tethering} filaments to the membrane \cite{mogilner2003force}), the zero-current boundary conditions no longer hold. More generally, the zero-current condition is characteristic of a nonequilibrium steady state, that is, one that is maintained through a constant input and subsequent dissipation of energy and for which a general theoretical formalism remains elusive \cite{evansblythe2002}.

\section*{Acknowledgements}

AJW acknowledges studentship funding from EPSRC under grant number EP/L015110/1.

\bibliography{RatchetSubmission}

\appendix

\section{Continuum limit of lattice Brownian ratchet} \label{app:ContLim}
We derive the diffusion equation \eqref{eq:DiffusionEquationBulk} and boundary conditions \eqref{eq:DiffusionEquationBoundary}, from the recurrence relations \eqref{eq:LatticeBulkRR} and \eqref{eq:LatticeBoundaryRR} that describe the lattice Brownian ratchet. 

\subsection{Diffusion equation}
With reference to Figure \ref{fig:LatticeRatchet}, define $a$ as a lattice spacing on this discrete system, such that $\vec{x} = (x_1,x_2,\dots x_N) = a\vec{i}$. With this included, the master equation \eqref{eq:LatticeBulkRR} becomes
\begin{widetext}
\begin{multline}
0 = -\left[2m+l+\sum_{n=1}^N(2q_n+r_n(x_n))\right]P(\vec{x}) + mP(\vec{x}-a\vec{1}) + (m+l)P(\vec{x}+a\vec{1}) \\
 + \sum_{n=1}^N q_nP(\vec{x}-a\vec{\hat{n}}) + \sum_{n=1}^N\left[q_n+r_n(x_n+a)\right]P(\vec{x}+a\vec{\hat{n}}) \;.
\end{multline}
Now, we treat $\vec{x}$ as a continuous vector and Taylor expand $P$ around $\vec{x}$ to second order. We find
\begin{multline}
0 \approx -\left[2m+l+\sum_{n=1}^N(2q_n+r_n(x_n))\right]P(\vec{x}) +  m\left(1 -a\sum_{n=1}^N\del{n}+\hf a^2 \sum_{n=1}^N\sum_{k=1}^N\del{n}\del{k}\right)P(\vec{x})  \nonumber \\
+ (m+l)\left(1 + a\sum_{n=1}^N\del{n}+\hf a^2 \sum_{n=1}^N\sum_{k=1}^N\del{n}\del{k}\right)P(\vec{x}) + \sum_{n=1}^N q_n\left(1-a\del{n}+\hf a^2 \del{n}^2\right)P(\vec{x}) \\
+ \sum_{n=1}^N \left(q_n+\left[1+a\del{n}+\hf a^2\del{n}^2\right]r_n(x_n)\right)\left(1+a\del{n}+\hf a^2 \del{n}^2\right)P(\vec{x}) 
\end{multline}
where we have used the shorthand $\del{n} \equiv \del{} / \del{} x_n$. This simplifies to
\begin{multline}
0 \approx  a\left(\sum_{n=1}^N\left[\del{n}r_n(x_n)+(r_n(x_n)+l)\del{n}\right]P(\vec{x})\right)   \\ 
 +  a^2\left(\left(m+\hf l\right)\sum_{n=1}^N\sum_{k=1}^N \del{n}\del{k}P(\vec{x})+\sum_{n=1}^N \left[\left(q_n+\hf r_n(x_n)\right)\del{n}^2+\hf\del{n}^2 r_n(x_n)\right]P(\vec{x}) \right)
\label{eq:diffeqnintermediate}
\end{multline}
since all $\mathcal{O}(a^0)$ terms cancel.
\end{widetext}

We now define a set of \emph{diffusion} and \emph{drift} rates, first for the membrane (subscript $M$) 
\begin{align}
\mu_M = l \;, \qquad \vec{\mu}_M \equiv \sum_{n=1}^N \mu_M\vec{\hat{n}} \;, \qquad D_M=am \; .
\end{align}
For the filaments, define
\begin{align}
\del{n}V(\vec{x}) = r_n \;, \qquad D_n=  aq_n
\end{align}
writing the drift $\del{n}V(\vec{x})$ in terms of a potential gradient. We then rewrite \eqref{eq:diffeqnintermediate}, retaining leading-order terms only:
\begin{multline}
0 = \sum_{n=1}^N\del{n}\Bigg[\left(\del{n}V(\vec{x})+\mu_M\right)P(\vec{x})  \\
{} + \left(D_M \sum_{k=1}^N \del{k}+ D_n\del{n}\right)P(\vec{x})\Bigg]
\end{multline}
which is the diffusion equation \eqref{eq:DiffusionEquationBulk}.

\subsection{Boundary conditions}
Starting from the master equation \eqref{eq:LatticeBoundaryRR} that applies when a filament is in contact with the membrane, we can follow a similar sequence of steps to obtain a boundary condition on the diffusion equation. This time we do not get full cancellation at $\mathcal{O}(a^0)$, so we need only expand to first order to obtain:
\begin{widetext}
\begin{multline}
0 \approx \Bigg[-\Bigg(m+q_k+\sum_{\substack{n=1 \\ n \neq k}}^N(2q_n + r_n(x_n))\Bigg)P(\vec{x}) + (m+l)\Bigg(1+a\sum_{n=1}^N\del{n}\Bigg)P(\vec{x}) \\
 + \sum_{\substack{n=1 \\ n \neq k}}^Nq_n\left(1-a\del{n}\right)P(\vec{x})  + \sum_{n=1}^N\left(q_n+\left[1+a\del{n}\right]r_n(x_n)\right)\left(1+a\del{n}\right)P(\vec{x})\Bigg]_{x_k=0}
\end{multline}
which simplifies to
\begin{equation}
0 \approx \Bigg[(l+r_k(x_k))P(\vec{x}) + a\left(q_k\del{k}+\sum_{n=1}^N \left[\del{n}r(x_n) +(m+l+r_n(x_n))\del{n}\right]\right)P(\vec{x})\Bigg]_{x_k=0} \;.
\end{equation}
\end{widetext}

Now, on using the above definitions of the drift and diffusion rates, we ultimately find
\begin{equation}
0 = \left[\left(\mu_M+\del{k}V_k(x_k) +D_k\del{k} + \sum_{n=1}^N D_M \del{n} \right)P(\vec{x})\right]_{x_k=0} \;.
\end{equation}
which, for all $k = 1,2,\dots N$ is the set of boundary conditions \eqref{eq:DiffusionEquationBoundary}. This is a first order equation, reflective of the deterministic dynamics on contact with the boundary. 

\subsection{Membrane velocity formula}
	\label{sec:MemVelFormula}
We now show in detail how to obtain Eq.~\eqref{eq:MembraneVelocityExpression}, the formula for the mean continuum membrane velocity, $v_M$, as a function of the various parameters in the system. We begin from a simple expression for the velocity in the discrete case, which we take a continuum limit of.

In the discrete system, the membrane will move at an average velocity $-l$ when no filaments are in contact with it, and at velocity $+m$ in any configuration $\vec{i}^{(c)}$ where one or more filaments are in contact (see Figure~\ref{fig:LatticeRatchet}):
\begin{align}
v_M &= -l (1-\mathcal{P}_{\rm contact}) + m\mathcal{P}_{\rm contact} \\
& = -l + (m+l)\sum_{\vec{i}^{(c)}}\mathcal{P}_{\vec{i}} \;.
\label{vmdef}
\end{align}
By convention, $v_M$ is positive if the membrane is moving to the right. Here, $\mathcal{P}_{\rm contact}$ is the overall probability of the membrane being in contact with any filament i.e.\ a sum over all configurations $\vec{i}^{(c)}$ where one or more filament contacts the membrane. With the parameters in Eq.~\eqref{eq:FilamentCLDefinitions}, we obtain from Eq.~\eqref{vmdef} in the continuum limit
\begin{widetext}
\begin{align}
v_M =& -l (1-\mathcal{P}_{\rm contact}) + m\mathcal{P}_{\rm contact} \\ 
 \approx& -l + (m+l)\sum_{n=1}^N \left[\prod_{\substack{m=1 \\ m\neq n}}^N\left(\sum_{m\geq0}\right) \mathcal{P}(\vec{i}|_{i_n=0})\right]  \\
\approx& -\mu_M + \left(\frac{D_M}{a}+\mu_M\right)\sum_{n=1}^N \left[\prod_{\substack{m=1 \\ m\neq n}}^N\left(\intt \frac{dx_m}{a}\right) a^N \mathcal{P}(\vec{x}|_{x_n=0})\right]\;. \label{eq:VelocityIntegralIntermediate}
\end{align}
\end{widetext}
where $\mathcal{P}(\vec{x}|_{x_n = 0})$ is the pdf evaluated at $x_n=0$. We have neglected any configurations where two or more filaments make contact: any such configurations would make an $\mathcal{O}(a)$ contribution to the velocity in Eq.~\eqref{eq:VelocityIntegralIntermediate}, as these terms will comprise fewer than $(N-1)$ integrals in $dx_m/a$. In the limit $a\to 0$, then, these terms will vanish. Taking this limit we recover Eq.~\eqref{eq:MembraneVelocityExpression}, 
\begin{equation*}
v_M =
-\mu_M + D_M\sum_{n=1}^N\left[\prod_{\substack{m=1 \\ m\neq n}}^N\left(\intt dx_m\right)\mathcal{P}(\vec{x}|_{x_n = 0})\right]\;.
\end{equation*}

\end{document}